\documentclass[conference]{IEEEtran}

\usepackage[english]{babel}
\usepackage{graphicx}
\usepackage{amssymb, amsmath, amsfonts, amsthm}
\usepackage{lipsum}
\usepackage{algorithm}
\usepackage{color,colortbl,multicol,multirow,hhline,array}
\usepackage{bm}
\usepackage{siunitx} 
\usepackage{enumerate}

\usepackage[printonlyused,nohyperlinks,nolist]{acronym}
\usepackage[capitalise]{cleveref}

\begin{acronym}[]
	\acro{1D}{one dimensional}
	\acro{2D}{two dimensional}
	\acro{3D}{three dimensional}
	\acro{3GPP}{3rd Generation Partnership Project}
	\acro{5G}{Fifth Generation}
	
	\acro{AoA}{Angle of Arrival}
	\acro{AAoA}{Azimuth Angle of Arrival}
	\acro{AoD}{Angle of Departure}
	\acro{AP}{Access Point}
	\acro{ARM}{Array-Response-Matched}
	\acro{ASA}{Azimuth spread of Arrival}
	\acro{ASD}{Azimuth spread of Departure}
	\acro{ASSQ}{Adaptive Search Space Quantization}
	\acro{AWGN}{Additive White Gaussian Noise}
	\acro{AWCSCGN}{Additive White Circular Symmetric Complex Gaussian Noise}
	\acro{ACF}{Auto-Correlation function}

	\acro{BD}{Block Diagonalization}
	\acro{BDIA}{Block Diagonalization - Interference Alignment}
	\acro{BLER}{Block Error Rate}
	\acro{BS}{Base Station}

	\acro{CDF}{Cumulative Distribution Function}
	\acro{CIR}{Channel Impulse Response}
	\acro{CMD}{Correlation Matrix Distance}
	\acro{CoMP}{Coordinated Multi-Point}
	\acro{CoF}{Cost of Feedback}
	\acro{CQI}{Channel Quality Indicator}
	\acro{CRONOS}{Cellular Radio Network Simulator}
	\acro{CSI}{Channel State Information}
	\acro{CSIT}{\ac{CSI} at the transmitter}

	\acro{DS}{Delay Spread}
	\acro{DD}{Doppler-Delay}
	\acro{DFT}{Discrete Fourier Transform}
	\acro{DoA}{Direction of Arrival}
	\acro{DoF}{Degrees of Freedom}
	\acro{DoP}{Dilution of Precision}
	
	\acro{EAoA}{Elevation Angle of Arrival}
	\acro{eNB}{evolved Node B}
	\acro{EZFB}{Effective Zero Forcing Beamforming}
	\acro{EsA}{Elevation spread of Arrival}
	\acro{EsD}{Elevation spread of Departure}

  \acro{FANTASTIC-5G}{Flexible Air iNTerfAce for Scalable service delivery wiThin wIreless Communication networks of the 5th Generation}
	\acro{FBR}{Feedback Rate}
	\acro{FDD}{Frequency-Division Duplex}
	\acro{FFT}{Fast Fourier Transform}
	\acro{FI}{Feedback Interval}
	\acro{FIR}{Finite Impulse Response}
	\acro{FLS}{Fully Loaded System}
	\acro{FR}{Frequency Response}
	\acro{FTP}{File Transfer Protocol}	
	\acro{FWHM}{Full Width at Half Maximum}

	\acro{GCS}{Global Coordinate System}
	\acro{GDoP}{Geometry \ac{DoP}}
	\acro{GF}{Geometry Factor}
	\acro{GNSS}{Global Navigation Satellite System}
	\acro{GoC}{Gain of CoMP}
	\acro{GoP}{Gain of Prediction}
	\acro{GPS}{Global Positioning System}
	\acro{GSCM}{Geometry-based Stochastic Channel Model}
	\acro{GSM}{Global System for Mobile Communications}

	\acro{HARQ}{Hybrid Automatic Repeat reQuest}
	\acro{HPBW}{Half Power Beam Width}
	\acro{HHI}{Heinrich Hertz Institute}	
	\acro{HetNet}{Heterogeneous Network}
	\acro{HRA}{High Resolution Algorithm}

	\acro{IA}{Interference Alignment}
	\acro{ICI}{Inter-Cluster Interference}
	\acro{ISD}{Inter Site Distance}
	\acro{ID}{Identity}
	\acro{i.i.d.}{independent and identically distributed}
	\acro{IDFT}{Inverse Discrete Fourier Transform}
	\acro{IFFT}{Inverse Fast Fourier Transform}

	\acro{JSDM}{Joint Spatial Division and Multiplexing}
	\acro{JT}{Joint Transmission}

	\acro{KF}{Ricean K-Factor}
	\acro{KPI}{key Performance Indicator}

	\acro{LBS}{Last Bounce Scatterer}
	\acro{LHCP}{Left Hand Circular Polarized}
	\acro{LoS}{Line of Sight}
	\acro{LSP}{Large Scale Parameter}
	\acro{LTE}{Long Term Evolution}
    \acro{LSF}{large-scale fading}

	\acro{MC}{Multi Cluster}
	\acro{MET}{Multi-user Eigenmode Transmission}
	\acro{MIMO}{Multiple-Input Multiple-Output}
	\acro{MISO}{Multiple-Input Single-Output}
	\acro{MMSE}{Minimum Mean Square Error}
	\acro{MPC}{Multi Path Component}
	\acro{MRT}{Maximum Ratio Transmission}
	\acro{MS}{Mobile Station}
	\acro{NMSE}{Normalized Mean Square Error}
	\acro{MSE}{Mean Square Error}
	\acro{MRP}{Maximum Received Power}
	\acro{MT}{Mobile Terminal}
	\acro{MU}{Multi-User}
	
	\acro{NGMN}{Next Generation Mobile Networks}
	\acro{NLoS} {Non Line Of Sight}
	\acro{NR}{New Radio}

	\acro{OFDM}{Orthogonal Frequency Division Multiplexing}

	\acro{PDoP}{Position \ac{DoP}}
	\acro{PG}{Path Gain}
	\acro{PL}{path Loss}
	\acro{PLS}{Partially Loaded System}
	\acro{PUCA}{Polarized Uniform Circular Array}

	\acro{QAM}{Quadrature Amplitude Modulation}
	\acro{QoE}{Quality of Experience}
	\acro{QuaDRiGa}{Quasi Deterministic Radio Channel Generator}
	\acro{QRDRLS}{QR-Decomposition Recursive Least Squares}

	\acro{RADAR}{Radio Detection and Ranging}
	\acro{RB}{Resource Block}
	\acro{RHCP}{Right Hand Circular Polarized}
	\acro{RLS}{Recursive Least-Squares}
	\acro{RLE}{Run-Length Encoding}
	\acro{RR}{Round Robin}
	\acro{RS}{Reference Signals}
	\acro{RSS}{Receive Signal Strength}
	\acro{RTT}{Round Trip Time}
	\acro{Rx}{Receiver}

	\acro{SAGE}{Space-Alternating Generalized Expectation-maximization}
	\acro{SCM}{Spatial Channel Model}
	\acro{SDMA}{Spatial Division Multiple Access}
	\acro{SF}{Shadow Fading}
	\acro{SVD}{Singular Value Decomposition}
	\acro{SNR}{Signal to Noise Ratio}
	\acro{SIR}{Signal to Interference Ratio}
	\acro{SINR}{Signal to Interference and Noise Ratio}
	\acro{SISO}{Single Input Single Output}
	\acro{SSF}{Small Scale Fading}
	\acro{STD}{Standard Deviation}
	\acro{SUS}{Semi-orthognal User Selection}
    \acro{SSF}{small-scale fading}
    \acro{SOS}{sum-of-sinusoids}

	\acro{TCP}{Transmission Control Protocol}
	\acro{TDD}{Time Division Duplex}
	\acro{TDoA}{Time Difference of Arrival}
	\acro{TOF}{Time of Flight}
	\acro{TP}{Throughput}
	\acro{TTI}{Time Transmission Interval}
	\acro{TUB}{Technische Universit\~{A}¤t Berlin}
	\acro{Tx}{Transmitter}

	\acro{UE}{User Equipment}
	\acro{UCA}{Uniform Circular Array}
	\acro{UDP}{User Datagram Protocol}
	\acro{ULA}{Uniform Linear Array}
	\acro{UML}{Unified Modeling Language}
	\acro{UMTS}{Universal Mobile Telecommunications System}
	\acro{UPA}{Uniform Planar Array}
	\acro{URA}{Uniform Rectangular Array}
	\acro{USA}{Uniform Square Array}
	\acro{UWB}{Ultra-Wideband}

	\acro{WGS}{World Geodetic System}
	\acro{WLAN}{Wireless Local Area Network}
	\acro{WINNER}{Wireless World Initiative for New Radio}
	\acro{WLS}{Weighted Least Square}
	\acro{WSS}{Wide-Sense Stationary}

	\acro{XPD}{Cross-Polarization Discrimination}
	\acro{XPR}{Cross Polarization Ratio}

	\acro{ZF}{Zero-Forcing}
	\acro{ZFPB}{Zero-Forcing Projection Based}

\end{acronym} 
\graphicspath{{figures/}}
\DeclareGraphicsExtensions{.pdf,.jpg,.png}

\def\BibTeX{{\rm B\kern-.05em{\sc i\kern-.025em b}\kern-.08em
    T\kern-.1667em\lower.7ex\hbox{E}\kern-.125emX}}

\begin{document}
\title{Evaluation of the Spatial Consistency Feature in the 3GPP GSCM Channel Model}
\author{
	\IEEEauthorblockN{Martin Kurras, Sida Dai, Stephan Jaeckel, Lars Thiele}
	\IEEEauthorblockA{Fraunhofer Heinrich Hertz Institute \\		
		Einsteinufer 37, 10587 Berlin, Germany \\
		Email: {martin.kurras@hhi.fraunhofer.de }} 		
}

\maketitle
\begin{abstract}
Since the development of 4G networks, \ac{MIMO} and later multiple-user \ac{MIMO} became a mature part to increase the spectral efficiency of mobile communication networks. An essential part of simultaneous multiple-user communication is the grouping of users with complementing channel properties. With the introduction of \ac{BS} with large amount of antenna ports, i.e. transceiver units, the focus in spatial precoding is moved from uniform to heterogeneous cell coverage with changing traffic demands throughout the cell and 3D beamforming. In order to deal with the increasing feedback requirement for \ac{FDD} systems, concepts for user clustering on second order statistics are suggested in both the scientific and standardization literature. Former \ac{3GPP} \ac{GSCM} channel models lack the required spatial correlation of small-scale fading. Since the latest release of \ac{3GPP} \acl{GSCM} this issue is claimed to be solved and hence our contribution is an evaluation of this spatial consistency feature.
\end{abstract}
\begin{IEEEkeywords}
GSCM, MIMO, Channel, Model, Spatial, Consistency
\end{IEEEkeywords}
\acresetall
\section{Introduction} \label{sec:introduction}
In mobile communications, the abstraction of the underlying wireless channel is an essential part. Parallel to increased computation and storage capacities, also the complexity of channel models has increased over the years \cite{ECS+98}, capturing more and more effects of real environments. Channel models can be categorized into statistical and deterministic channel models \cite{ECS+98}. A combination of the statistical and deterministic modeling components can be found in the so-called \acp{GSCM}. This type of model has been used as a basis for the standardization of the fourth generation (4G) of the mobile network. One important requirement for this work were standardized channel models that can be used to evaluate and compare the different proposals against each other. This was provided with the \ac{3GPP}-\ac{SCM} in 2003 \cite{3GPP12-25996}. Since then, the family of \acp{GSCM} is under constant development, evolving to enable performance evaluation of new techniques such as ``massive'' \ac{MIMO} \cite{RPL+13}, which led to the \ac{3D} extension of \ac{GSCM} \cite{3GPP17-36873} or the extension to frequencies above \si{6}{ GHz} \cite{3GPP17-38901}.

Another recently added feature is ``spatial consistency'' \cite{3GPP17-38901}, because one major drawback of the \acp{GSCM} channel has been the lack of realistic correlation in the \ac{SSF} \cite{BWW+17}. While \acp{LSP}, such as \ac{AoA} spread, \ac{AoD} spread, delay spread, K-factor, and \ac{SF}, are correlated both in space and with each other to ensure consistent channel properties of closely located mobile users, \ac{SSF} has been uncorrelated \cite{3GPP12-25996}. As a consequence, the signal paths of users located next to each other had a correlated angular spread (\ac{AoA}), but their individual directions were independently generated. These uncorrelated \ac{MPC} directions are caused by ``random'' generation of positions of the scattering clusters. This behavior intuitively contradicts experienced causality in reality and observations in channel measurements \cite{GERT15}.

Uncorrelated \ac{SSF} was also the reason why massive \ac{MIMO} schemes that rely on spatial consistency could not be reliably investigated with \acp{GSCM} before. One promising massive \ac{MIMO} schemes is \ac{JSDM} \cite{ANAC13} which is based on the assumption that users can be clustered into groups with ``similar'' covariance matrices \cite{NAAC14,KFT15}. What ``similar'' covariance matrices exactly means and how it is measured is described in \cref{sec:kpis}. However, in publicly available evaluations of the spatial consistency feature, metrics that measure the similarity of \ac{SSF} are missing, e.g. in \cite{3GPP17-38901} (further reference to R1-1700990) the coupling loss,  wideband \ac{SINR}, cross-correlation of delays, \acp{AoA}, and \ac{LoS}/\ac{NLoS} status are investigated, in \cite{TKC16} the impact on the signal to interference ratio is studied, and in \cite{JRB+17} the path power, delay, \ac{AoA} and \ac{AoD} of a single user is provided.

In contrast to previous work, this paper evaluates spatial consistency in terms of \ac{AoA} difference and covariance matrix similarity between users. The investigated two dimensional parameter space is the distance between the two users and the spatial decorrelation distance denoted by \(d_\lambda\), a parameter that adjusts the degree of ``correlation'' between the positions of the scattering clusters observed from distinct user positions. For evaluation we use the open source available \ac{QuaDRiGa} channel model \cite{JRB+17}, where the spatial consistency feature is integrated since version 2.0 and which is calibrated against the \ac{3GPP} \ac{NR} channel model specification \cite{3GPP17-38901}. The details of the implementation are available in the source code at \cite{JRB+17}. 

The remainder of this paper is organized as follows. In \cref{sec:system_model} the principles of the \ac{3GPP} \ac{GSCM} channel model are introduced. \cref{sec:pos_scat_cluster} describes the positioning of the scattering clusters, with and without the spatial consistency feature. Following, \cref{sec:kpis} defines the performance metric used for numerical evaluation in \cref{sec:numerical_results}. Finally, \cref{sec:conclusion} concludes this paper.
\section{Channel Model}\label{sec:system_model}
The wireless channel between a transmitter and a receiver in the frequency domain is denoted as \(\mathbf{H} \in \mathbb{C}^{n_r \times n_t}\), where \(n_r\) and \(n_t\) are the number of antennas at the receiver and the transmitter, respectively. \(\mathbf{H}_n\), where \(n\) is the sample frequency, can be decomposed as
\begin{equation}
	\mathbf{H}_n = \sum_{l=1}^{L} \mathbf{G}_l \cdot e^{-2\pi j \cdot f_n \cdot \tau_l},
	\label{eq:channel_mtx_freq}
\end{equation}
where the index \(l = \left[1, \ldots, L \right]\) denotes the path number and \(\mathbf{G}_l\) is the \ac{MIMO} coefficient matrix. \(f_n\) is the \(n\)-th sample frequency in [Hz] relative to the beginning of the used  bandwidth and \(\tau_l\) is the delay of the \(l\)-th path in seconds. The coefficient matrix \(\mathbf{G}_l\) has \(n_r\) rows and \(n_t\) columns. It contains one complex-valued channel coefficient for each antenna pair. 

One major advantage of \acp{GSCM} is that they allow the separations of propagation and antenna effects. Therefore, it is essential to have a description of the antenna that captures all relevant effects that are needed to accurately calculate the channel coefficients in the model. Antennas do not radiate equally in all directions. Hence, the radiated power is a function of the angle. Consequently, the individual values of the coefficients are a result of the attenuation of a path, the weighting by the antenna radiation patterns, and the polarization.

Thus, the complex-valued amplitude $g_{r,t,l}$ of the \(l\)-th path between the \(t\)-th transmit and \(r\)-th receive antenna is given by
\begin{equation}\label{eq:single_channel_coeff_simple}
	g_{r,t,l} = \sqrt{P_l} \cdot \mathbf{F}_{r}(\phi^a_l,\theta^a_l)^T \cdot \mathbf{M} \cdot \mathbf{F}_{t}(\phi^d_l,\theta^d_l) \cdot e^{-j \frac{2\pi}{\lambda}\cdot d}\text{,}
\end{equation}
where $\mathbf{F}_r$ and $\mathbf{F}_t$ describe the polarimetric antenna response at the receiver and the transmitter, respectively, \(r = \left[ 1, \ldots, n_r\right]\) is the receive antenna index and \(t = \left[1, \ldots, n_t \right]\) is the transmit antenna index.
$P_l$ is the power of the \(l\)-th path, $\lambda$ is the wavelength, $d$ is the length of the path, $(\phi^a_l,\theta^a_l)$ are the arrival and $(\phi^d_l,\theta^d_l)$ the departure angles that are defined by the position of the receiver, transmitter, and scattering clusters. $\mathbf{M}$ is the $2 \times 2$ polarization coupling matrix. This matrix describes how the polarization changes on the way from the transmitter to the receiver. The \ac{XPR} quantifies the separation between two polarized channels due to different polarization orientations. $\mathbf{M}$ is then often modeled by using random coefficients ($Z_{\theta\theta}$, $Z_{\theta\phi}$, $Z_{\phi\theta}$, $Z_{\phi\phi}$) as
\begin{equation}\label{eq:random_pol_coupling}
	\mathbf{M} = \left(
		\begin{array}{cc}
			Z_{\theta\theta} & \sqrt{1/\mathrm{XPR}} \cdot Z_{\theta\phi} \\
			\sqrt{1/\mathrm{XPR}} \cdot Z_{\phi\theta} & Z_{\phi\phi} \\
		\end{array}
	\right)\text{,}
\end{equation}
where $Z \sim \exp \left\{ j \cdot \mathcal{U}(-\pi,\pi) \right\}$ introduces a random phase.

Of all the different ways to describe polarization \cite{GMP07}, the polar spherical polarization basis is the most practical for \acp{GSCM}. In the polar spherical basis, the antenna coordinate system has two angles and two poles. The elevation angle \(\theta\) is measured relative to the pole axis. A complete circle will go through each of the two poles, similar to the longitude coordinate in the \ac{WGS}. The azimuth angle \(\phi\) moves around the pole, similar to the latitude in \ac{WGS}. Thus, the antenna is defined in \emph{geographic} coordinates, the same coordinate system that is used in the channel model. Hence, deriving the antenna response from the previously calculated departure and arrival angles is straightforward. The electric field is resolved onto three vectors which are aligned to each of the three spherical unit vectors $\mathbf{\hat{e}}_{\theta}$, $\mathbf{\hat{e}}_{\phi}$ and $\mathbf{\hat{e}}_{r}$ of the coordinate system. In this representation, $\mathbf{\hat{e}}_{r}$ is aligned with the propagation direction of a path. In the far-field of an antenna, there is no field in this direction. Thus, the radiation pattern consists of two components, one is aligned with $\mathbf{\hat{e}}_{\theta}$ and another is aligned with $\mathbf{\hat{e}}_{\phi}$. The polarimetric antenna responses $\mathbf{F}_r$ and $\mathbf{F}_t$ are each described by by a 2-element vector
\begin{equation}\label{eq:pattern_response}
  \mathbf{F}(\theta,\phi) = \left(
    \begin{array}{c}
      F^{[\theta]} (\theta,\phi) \\
      F^{[\phi]} (\theta,\phi)\\
    \end{array}
  \right)\text{.}
\end{equation}
Further details on the channel model can be found in \cite{Jae17}.

\section{Spatial Consistency}\label{sec:pos_scat_cluster}
\acp{GSCM} consist of two main components: a stochastic part that generates a random propagation environment, and a deterministic part that lets transmitters and receivers interact with this environment. In order to make realistic predictions of the wireless system performance, the random environment must fulfill certain statistical properties which are determined by measurements. Then, for a given set of model parameters, the joint spatial correlation of these parameters must be captured for a large number of transceivers. This is done in the so-called \ac{LSF} model. A subsequent \ac{SSF} model generates individual scattering clusters for each \ac{MT}.

\acp{LSP} do not change rapidly. Typically, they are relatively constant for several meters. An example is the \ac{SF} which is caused by buildings or trees blocking a significant part of the signal. The so-called decorrelation distance of the \ac{SF}, \emph{i.e.}, the distance a \ac{MT} must move to experience a significant change in the \ac{SF}, is in the same order of magnitude as the size of the objects causing it. Thus, if a \ac{MT} travels along a trajectory or if multiple \acp{MT} are closely spaced together, their \acp{LSP} are correlated. A common approach to model such correlation is by filtered Gaussian-distributed random numbers \cite{Bakowski2011}. However, when it comes to spatial consistency, the positions of the scattering clusters must also be spatially correlated. A modeling approach for this has been introduced by the 3GPP new-radio model \cite{3GPP17-38901} where it is suggested that ``spatially consistent powers/delays/angles of clusters are generated". This requires that all random variables that determine the location of the scattering clusters are spatially correlated. For a moderate scenario with 12 clusters and 20 sub-paths per cluster, this would result in 2232 random variables. Compared to the 7 variables needed for the \ac{LSF} model, the filtering approach would require prohibitively large amounts of memory.

In order to reduce the memory requirements, the distance-dependent correlation of the \ac{SSF} parameters can be modeled by means of the \ac{SOS} method. The method for generating these correlations is adopted from \cite{Wang2005b}. The main advantage of using the \ac{SOS} method lies in the low memory requirements since it only requires a relatively small number of sinusoids coefficients to generate spatially correlated parameters for a large volume of space. Spatially correlated random variables $k(x,y,z)$ generated by the SOS method are Normal-distributed.
\begin{equation}
    k(x,y,z) = \mathcal{N}(\mu,\sigma^2)
\end{equation}
where $\mathcal{N}(\mu,\sigma^2)$ denotes a Normal distribution with mean $\mu$ and standard deviation $\sigma$. Other distributions can be generated by a mapping operation. The parameter value $k$ is a function of the \ac{MT} location in \ac{3D} Cartesian coordinates $(x,y,z)$. The \ac{1D} spatial \ac{ACF} describes how fast the local mean of $k(x,y,z)$ evolves as the \ac{MT} moves. The \ac{ACF} is usually modeled as an exponential decay function
\begin{equation}\label{eq:exponential_corr_decay_map}
    \rho(d) = \exp\left(-\frac{d}{d_\lambda}\right)\text{,}
\end{equation}
with $d$ as the distance between two positions and $d_\lambda$ as the so-called decorrelation distance, i.e.\ the distance at which the correlation drops to $e^{-1}\approx 0.37$ \cite{Gudmundson1991_Correlation}. The \ac{SOS} method now approximates $k(x,y,z)$ as
\begin{align}
	\label{eq:sos_3D_process}
	\hat{k}(x,y,z) = \sum_{n=1}^N a_n \cos\{ 2\pi (f_{x,n} x + \nonumber \\
	f_{y,n} y + f_{z,n} z ) + \psi_n \}
\end{align}
with $N$ sinusoids. The variables $a_n$, $f_n$, and $\psi_n$ denote the amplitude, the frequency, and the phase of a sinusoid, respectively. The amplitudes $a_n$ and the frequencies $f_n$ are determined in a way that $\hat{k}(x,y,z)$ has the same approximate \ac{ACF} as $k(x,y,z)$ and the \ac{CDF} is close to Gaussian density if $N$ is sufficiently large. The phases $\psi_n$ are random variables distributed in the range from $-\pi$ to $\pi$. Hence, exchanging the $\psi_n$ while keeping $a_n$ and $f_n$ fixed creates a new set of spatially correlated random variables at minimal computational cost. The main challenge of using the \ac{SOS} method is to find the frequencies $f_{x,n}$, $f_{y,n}$, and $f_{z,n}$ for a given \ac{ACF}. Suitable methods for doing this have been introduced by \cite{Patzold1996} which have been adopted for the implementation of the spatial consistency feature.
\section{Performance Metrics} \label{sec:kpis}
In this section, we introduce three performance metrics that are considered representative for evaluating the spatial consistency feature. These metrics are based on either angular distance or the comparison between the covariance matrices of the users. 
\subsection{Angular Distance}
In wireless channels, spatial consistency is shown in three aspects: 1) Large-scale parameters, 2) \ac{LoS}/\ac{NLoS} state of a link and 3) Distribution of scattering clusters according to user positions. Since the large-scale parameters are always spatially consistent, we focus only on the \ac{SSF} parameters in this paper. For example, two closely spaced receivers will not only experience similar \ac{SF} and angular spread, they should also be able to see the same clusters. In the \ac{GSCM}, the exact positions of each scattering cluster are calculated. Moreover, the \ac{AAoA} $\phi$ and \ac{EAoA} $\theta$ of each path are known at each user. Therefore, the arrival angles are the direct representations of each \ac{MPC} from the user's perspective. Here, in order to examine the similarity of the \ac{SSF} parameters, we consider the absolute angular difference between two receivers as the angular distance. Given the \acp{AAoA} and \acp{EAoA} of receiver $i$ and $j$,  the angular distance for \ac{MPC} $l$ is defined by
\begin{subequations}\label{eq:angular_distance}
	\begin{align}
       &\Delta\phi_{l}=\left\{\begin{array}{l}
         |\phi_{i,l}-\phi_{j,l}|,\quad\textrm{if }|\phi_{i,l}-\phi_{j,l}|<\pi\\
		  2\pi-|\phi_{i,l}-\phi_{j,l}|,\quad\textrm{otherwise}
		  \end{array}\right.\label{eq:angular_distance_a}\\
		&\Delta\theta_{l}=|\theta_{i,l}-\theta_{j,l}|\label{eq:angular_distance_b},
	\end{align}
\end{subequations}
where \cref{eq:angular_distance_a} always selects the value smaller than $\pi$ as the angular distance due to the circular nature of \ac{AAoA}. This nature is not observed in \ac{EAoA} since the \ac{EAoA} value is restricted to $[-\pi/2,\pi/2]$. It is evident to see that the angular distance is lower bounded by 0 and  a small value of angular distance indicates high similarity in the \ac{SSF} parameters and therefore a high spatial correlation. A more general way to study the spatial consistency regardless of the number of \acp{MPC} involved is to average all angular distances over existing \acp{MPC}. \cref{sec:numerical_results} shows this average angular distance as a function of distance between users. The average angular distance is therefore given by
\begin{equation}\label{eq:avg_angular_distace}
	\Delta\phi=\frac{1}{L}\sum_{l}\Delta\phi_{l}\quad\mathrm{and}\quad\Delta\theta=\frac{1}{L}\sum_{l}\Delta\theta_{l},
\end{equation}
where $L$ is the total number of \acp{MPC}.
\subsection{Covariance Matrix}\label{sec:cov_mtx}
The authors in \cite{ANAC13} proposed \ac{JSDM}, a novel way to enable massive \ac{MIMO} gains in \ac{FDD} systems. With this approach the training overhead by partitioning users into groups is significantly decreased. All users within the same group share similar second-order statistics (e.g. in the form of covariance matrices). Therefore the channel covariance has been extensively researched to study the spatial correlation between users. With the given channel coefficients from \cref{eq:channel_mtx_freq}, we are able to obtain the covariance matrix $\mathbf{R}\in\mathbb{C}^{n_t\times n_t}$ by the following equation
\begin{equation}\label{eq:covariance_matrix}
	\mathbf{R}=\mathbb{E}[\mathbf{H}^\mathrm{H}\mathbf{H}].
\end{equation}
The covariance matrix is assumed to be slow-varying in time and depends on user mobility. The idea of grouping receivers with sufficiently similar covariance matrices inspired several similarity measures, some of which are also suitable for measuring the spatial consistency feature. Motivated by this, we investigate the two performance metrics based on covariance matrices in this subsection. The first makes use of the chordal distance criterion, the second adapts the \ac{CMD} measure introduced in \cite{MHA+17}.
\subsubsection{Chordal distance}\label{sec:chordal_dist}
The chordal distance has been introduced by \cite{GVL12} and is widely used as a distance metric between subspaces. The authors in \cite{MSEA03} used chordal distance as a metric for packings in Grassmannian spaces. Moreover, \cite{NAAC14, KFT15} came up with different user grouping algorithms for \ac{JSDM}, with chordal distance being the distance criterion. The idea of grouping closely placed users by studying the similarity between channel covariance also suits our intention well. Hence we utilize chordal distance as our second evaluation metric for spatial consistency. Given the covariance matrices of two users ($\mathbf{R}_1$,$\mathbf{R}_2$), the chordal distance is expressed by
\begin{equation}
	\label{eq:chordal_distance}
	d_\mathrm{C}(\mathbf{R}_1,\mathbf{R}_2)=\|\mathbf{R}_1\mathbf{R}_1^\mathrm{H}-\mathbf{R}_2\mathbf{R}_2^\mathrm{H}\|^2_\mathrm{F},
\end{equation}
where  \(\| \cdot \| _\mathrm{F}\) denotes the Frobenius norm. When studying the spatial correlation of two users, we should expect the chordal distance to decrease as the distance between the users drops. In fact, we consider two users are strongly spatially correlated if the chordal distance is smaller than a selected threshold $\epsilon_\mathrm{C}$.
\subsubsection{Correlation Matrix Distance}\label{sec:cmd}
The \ac{CMD} proposed in \cite{HCOB05} was a novel measure to track the changes in spatial structure of non-stationary \ac{MIMO} channels. The same metric was used in \cite{MHA+17} as a covariance matrix measure for user clustering for \ac{JSDM}. In this way, it is possible to lower the complexity of predicting the threshold involved in the chordal distance based clustering algorithms. According to \cite{MHA+17}, the similarity measure based on \ac{CMD} is expressed as
\begin{align}
	\label{eq:cmd}
	d_\mathrm{CMD}(\mathbf{R}_1,\mathbf{R}_2) & = 1-\mathrm{CMD} \left(\mathbf{R}_1,\mathbf{R}_2 \right) \nonumber \\
	& =\frac{\mathrm{Tr}(\mathbf{R}_1^\mathrm{H}\mathbf{R}_2)}{\|\mathbf{R}_1\|_\mathrm{F}\cdot\|\mathbf{R}_2\|_\mathrm{F}},
\end{align}
where \(\mathrm{Tr}\) denotes the ``trace'' operator as the sum of the diagonal elements. 
This similarity measure is $1$ in the case of $\mathbf{R}_1$ and $\mathbf{R}_2$ being collinear, $0$ in the case of $\mathbf{R}_1$ and $\mathbf{R}_2$ being orthogonal. Therefore, with the spatial consistency feature, $d_\mathrm{CMD}$ between two spatially correlated users should have a value close to $1$. With a properly defined threshold value $\epsilon_\mathrm{CMD}$, we are able to find the distance within which the users are considered strongly spatially correlated.

\section{Numerical Results} \label{sec:numerical_results}
For the evaluation of the spatial consistency feature we used \ac{QuaDRiGa} version 2.0 which implements the 3GPP new radio model \cite{3GPP17-38901} together with the \ac{SOS}-based spatial consistency model. The center frequency was set to \si{2} {GHz} in an urban macro \ac{NLoS} scenario, where the number of modeled \acp{MPC} is \(L=5\). A summary of the simulation parameters is listed in \cref{tab:sim_assump}.

\begin{table}
	\centering
		\begin{tabular}{m{3.5cm}|m{3.5cm}}
			Parameter 								& Value \\ \hline \hline
			\acs{QuaDRiGa} Version		& 2.0.0 \\ \hline
			Center frequency 					& \si{2} {GHz} \\ \hline
			Scenario									& UMa NLoS, \acs{3GPP} 38.901 \cite{3GPP17-38901} \\ \hline
			Number of \acsp{MPC} \(L\)& 5 \\ \hline
			Number \acs{BS}	antennas	&	64 \\ \hline
			Distribution \acs{BS} antennas	& \(8 \times 8\) \acs{UPA} \\ \hline
			\acs{BS} antenna pattern				& \(65^\circ\) \acs{HPBW} in azimuth and elevation \\ \hline
			Number \acs{MS}	antennas				&	1 \\ \hline
			\acs{MS} antenna pattern				& Isotropic  \\ \hline
			Bandwidth frequency channel			& \si{18}{MHz} \\ \hline
			Number of \acs{OFDM} subcarriers & 100 \\
		\end{tabular}
	\caption{Simulation assumptions}
	\label{tab:sim_assump}
\end{table}

\begin{figure}
	\centering
		\includegraphics[width=\linewidth]{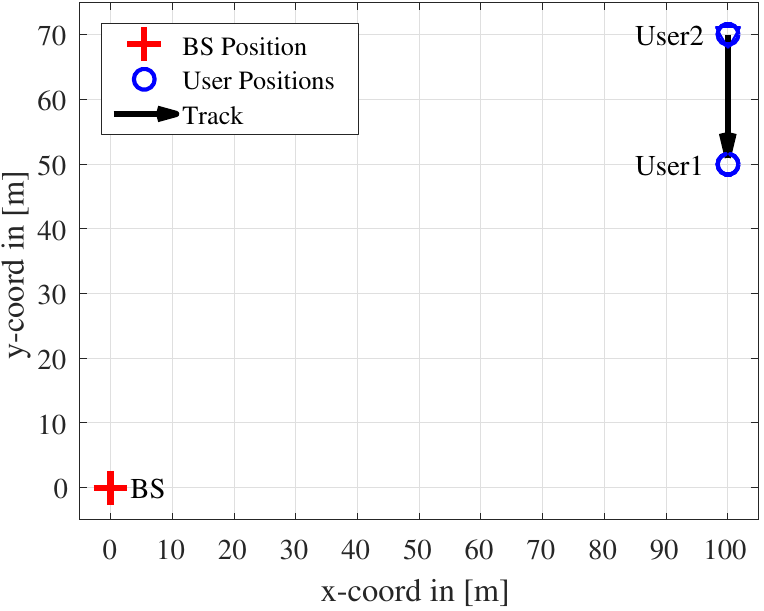}
	\caption{Deployment of \acs{BS} and users.}
	\label{fig:01_layout_visualize_adapt}
\end{figure}

A scenario with two users separated by a distance of \si{20}{ m} is considered, see \cref{fig:01_layout_visualize_adapt}. Therein, user two moves straight towards user one, labeled as ``Track'' in \cref{fig:01_layout_visualize_adapt}. The distance between user one and user two is denoted as \(d^{(\mathrm{1,2})}\). We assume downlink transmission from the \ac{BS} to the users such that the \ac{AoA} means angles observed by the users.

\begin{figure}
	\centering
	\includegraphics[width=\linewidth]{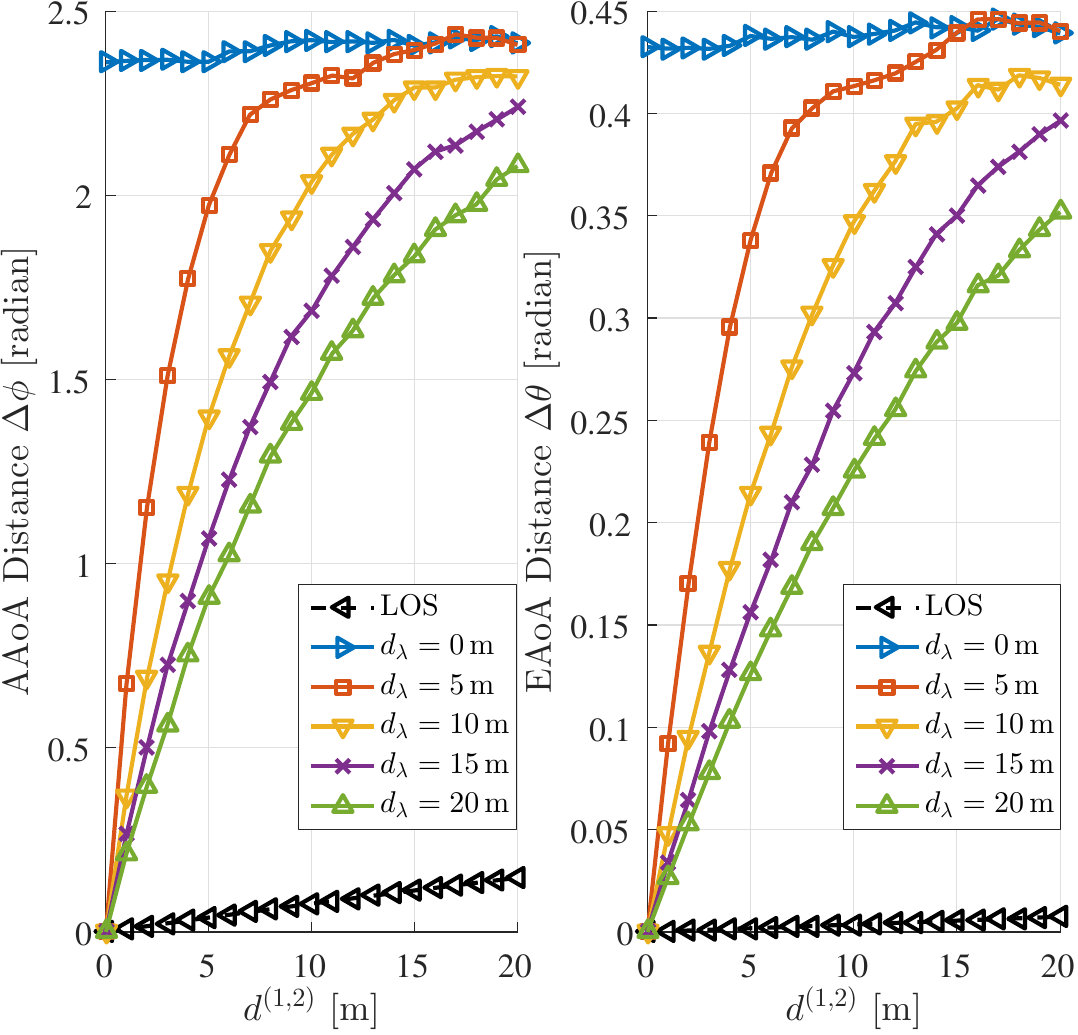}
	\caption{\acsp{MPC} angular distance according to \cref{eq:avg_angular_distace} over the distance between user 1 and user 2 , azimuth \acs{AoA} on left hand side and elevation \acs{AoA} on right hand side.}
	\label{fig:02_difference_aoa_eoa_adapt}
\end{figure}

First, the \ac{AoA} differences \(\Delta\phi\) and \(\Delta\theta\) according to \cref{eq:avg_angular_distace} are shown in \cref{fig:02_difference_aoa_eoa_adapt}, the \ac{AAoA} on the left hand side and the \ac{EAoA} on the right hand side. It can be observed that depending on the the spatial decorrelation distance \(d_\lambda\) the slope of angular distance decreases at a certain point, e.g. for \(d_\lambda=5\)\,m at \(\approx 7\)\,m. Note, that in \cref{fig:02_difference_aoa_eoa_adapt} the mean over the \(L=5\) \acp{MPC} is given in [radian]. The difference in the value range of the \ac{AAoA} and \ac{EAoA} in case of uncorrelated scattering positions \(d_\lambda=0\)\,m is caused by the``urban macro'' scenario assumption, where in measurements \acp{BS} are on the roof top of buildings limiting the elevation angular spread which is smaller than the azimuth angular spread.

\begin{figure}
	\centering
		\includegraphics[width=\linewidth]{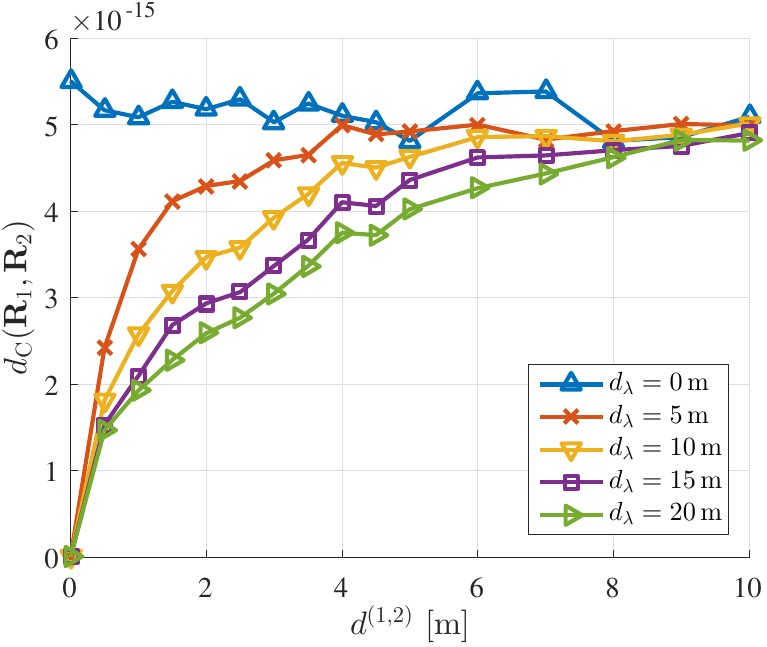}
	\caption{Chordal distance between user covariance matrices according to \cref{eq:chordal_distance} over the distance between user 1 and user 2.}
	\label{fig:03_chordal_distance_adapt}
\end{figure}

\cref{fig:03_chordal_distance_adapt} shows the Chordal distance \(d_\mathrm{C}\left(\mathbf{R}_1 \mathbf{R}_2\right)\) according to \cref{eq:chordal_distance} between user one and two. Therein, two observation are worthwhile to mention. First, the chordal distance is not a normalized measure and the value range, which is \( < 6 \cdot 10^-15\) shown in \cref{fig:03_chordal_distance_adapt}, depends on the power of the channel coefficients. Thus, any threshold \(\epsilon_\mathrm{C}\) for similarity between two users depends on the \acp{LSP} and path-loss of the two users. Second, the chordal distance \(d_\mathrm{C}\) decreases to zero as the distance drops, superposed by a \(d_\lambda\) depended threshold, where the slope of the decrease changes significantly, e.g. for \(d_\lambda = 20\)\,m at approximately 5\,m.

\begin{figure}
	\centering
		\includegraphics[width=\linewidth]{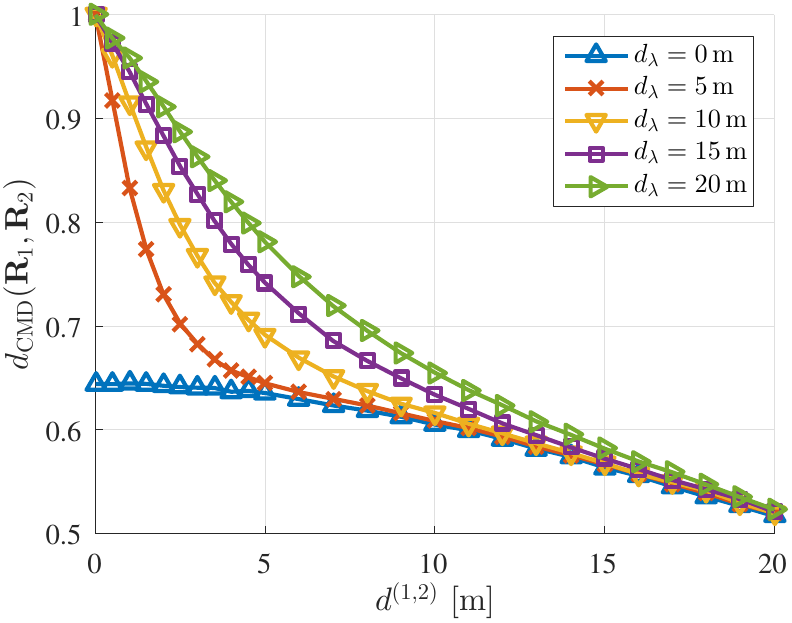}
	\caption{\(d_\mathrm{CMD}\) between covariance matrices according to \cref{eq:cmd} over the distance between user 1 and user 2.}
	\label{fig:03_cmd_adapt}
\end{figure}

Finally, \cref{fig:03_cmd_adapt} shows the \(d_\mathrm{CMD}\) according to \cref{eq:cmd} over the distance between user one and two. In contrast to the chordal distance, the \(d_\mathrm{CMD}\) is a normalized metric between zero and one, where one is reached for collinear covariance matrices, see \cref{sec:cmd}. Thus, large scale parameter independent thresholds can be defined, e.g. in \cite{MHA+17} a similarity threshold of \(\epsilon_\mathrm{CMD} = 0.95\) is set as a condition for users to be in the same cluster. In our numerical evaluation such a value is approximately achieved at \si{1}{m} distance for \(d_\lambda \geq15\)\,m. On the other hand, even for \(d_\lambda=0\)\,m (e.g. spatial consistency feature disabled), a decreasing \(d_\mathrm{CMD}\) is observed in \cref{fig:03_cmd_adapt} with a saturation at \(\lim_{d^{(\mathrm{1,2})} \rightarrow 0\mathrm{m}} d_\mathrm{CMD}\left(d_\lambda = 0\,\mathrm{m} \right) \approx 0.64\).
This is caused by the correlation of the \ac{LSP} in space, see \cref{sec:pos_scat_cluster}. The difference of the \ac{AoA} and \ac{AoD} spreads, delays, coupling loss, K-factor, and \ac{SF} also decreases with the distance and thus increases the \(d_\mathrm{CMD}\) even for \(d_\lambda = 0\)\,m.

\section{Conclusion} \label{sec:conclusion}
In this work, the recently proposed spatial consistency feature is evaluated. Thereby, several performance metrics are considered, showing that the spatial consistency feature works as expected. In a scenario where one user drifts to another, the angle distance of observed scattering clusters and chordal distance of the covariance matrices converge to zero, whereas the \(d_\mathrm{CMD}\) approaches one. Results also show the impact of the ``decorrelation distance'' parameter \(d_\lambda\), which tunes the degree of correlation between scattering cluster positions. The larger the \(d_\lambda\) the stronger the correlation.

The results in the paper enable researchers to parametrize the \ac{QuaDRiGa} channel model for simulations that require spatial consistent \ac{SSF}.  Furthermore, this work confirms that the \ac{3GPP} \ac{GSCM} can be used for reliable research and performance evaluations of \ac{JSDM} or similar user clustering based schemes.

\bibliographystyle{IEEEtran}
\bibliography{library_spatial_consistency}
\end{document}